\documentclass[runningheads]{llncs}
\usepackage[T1]{fontenc}
\usepackage{graphicx}
\begin{document}
\title{The Missing Layer: Why EdTech Needs Design-Time Generative UI, Not Just Runtime Personalization}
\author{Seyed Parsa Neshaei \and
Abhinand Shibu \and
Fatma Betül Güres}
\authorrunning{S. P. Neshaei et al.}
\institute{EPFL, Lausanne, Switzerland\\
\email{\{seyed.neshaei, abhinand.shibu, fatma-betul.gures\}@epfl.ch}}
\maketitle              % typeset the header of the contribution
\begin{abstract}
% AI
The dominant paradigm in using generative UI (GenUI) for adaptive EdTech considers the use of AI as a runtime engine: content is authored once in a fixed form, and AI adapts delivery dynamically based on learner needs, behaviors, or profiles. We argue that this paradigm has an issue: it moves the burden of accessibility and representation diversity onto systems that see learners only after content has already been locked into particular details. For learners who might need audio-first, simplified text, interactive, or low-bandwidth representations, runtime adaptation is too late and too costly to be equitable at scale, and might lead to inaccurate learning content due to the inability to conduct verification at scale. We propose an alternative method: accessibility belongs in the authoring layer. Specifically, we advocate for a card-based GenUI paradigm, in which educational content is encoded as modality-agnostic semantic units, and GenAI produces multiple interface representations, such as interactive, audio, text-simplified, or low-bandwidth, at learning design time to be verified by the instructor before it reaches any learner. This shifts the AI intervention from delivery to creation, embeds Universal Design for Learning principles into the authoring workflow, and removed per-learner inference costs. We situate this idea against recent work on GenUI, multimodal content generation, adaptive authoring, and equitable delivery, and argue that realizing this goal requires closer integration of AI, HCI, and learning sciences than what either of those communities has so far provided.

\end{abstract}
\vspace{-4pt}
\section{Introduction}
\vspace{-4pt}

% ai
Generative UI (GenUI), the use of large language models (LLMs) to produce interface components dynamically rather than rendering pre-designed layouts, has recently emerged as a promising direction for HCI and EdTech \cite{tu2026maic,pott2026taxonomy}. Rather than presenting a fixed interface to all learners, GenUI systems can generate interactive components, explanations, or visualizations on the fly, tailored to context, and in the case of EdTech, to each students' own behavior and profile. Early educational applications have shown encouraging results: MAIC-UI \cite{tu2026maic}, for instance, generates interactive STEM courseware from instructor-uploaded documents, and the study finds gains in learning agency in classroom deployment.

However, many applications of GenUI have focused at \textit{delivery} time \cite{pott2026taxonomy}: a learner interacts with a system, and the system generates or adapts interface components in response to that interaction. This is a natural framing, as it mirrors the broader paradigm of runtime personalization that has dominated the adaptive learning line of research for decades \cite{maity2024generative}; content is authored in a fixed form, and the intelligence (AI) operates on top of it at the moment of delivery.

We argue that this framing misses a prior and more consequential intervention point. Runtime GenUI operates on content that has already been fixed in a particular modality. For learners who require audio-first, low-bandwidth, or cognitively simplified representations, this is a fundamental limitation rather than an engineering gap: the content was never structured to support those representations in the first place. Moreover, generating diverse representations at runtime will lead to per-learner inference costs, particularly limiting its usage in under-resourced or low-connectivity deployment contexts \cite{cannanure2026teacher}. Also, when representations are generated at delivery time without prior review, content accuracy cannot be guaranteed; the instructor who authored the material and understands its domain is no longer in the loop at the moment generation occurs \cite{denny2023can}.
% We argue that GenAI makes such a mechanism possible now, but only if it is applied at the authoring layer rather than being deferred to runtime.

In this paper, we propose applying GenUI at the authoring layer of educational material. Specifically, loosely inspired by the idea behind HyperCard \cite{bowers1990hypercard}, we advocate for organizing educational content around cards: bounded modality-agnostic semantic units that encode a learning objective independently of any particular interface representation. At authoring time, a GenUI engine produces a defined set of representations from each card, e.g., interactive, audio, simplified text, and low-bandwidth plain text, which the instructor can review and correct before any learner sees them. The resulting cards can be added to card repositories, allowing deployability across different contexts using clients showing for each card the UI version appropriate for the specific learner.

The card boundary is central to the tractability of this approach. Generating representations for an unconstrained content space at runtime is both computationally expensive and pedagogically unsafe, as the space of possible learner contexts and modalities is unbounded. Constraining generation to the defined format of a card, with explicitly defined learning objectives and educational content, makes the generation problem tractable and keeps a domain-knowledgeable human in the loop before content reaches learners.

The progress to realize and evaluate this paradigm raises open questions for AI, HCI, and learning sciences that no single community might be able to address alone. On the AI side, generating representations that are pedagogically equivalent across modalities requires understanding going beyond current multimodal generation benchmarks \cite{bi2026eduillustrate}. Also, designing authoring interfaces that make multi-representation review seamless and effortless for non-technical educators is an open HCI problem, with we suggest to be built upon prior work in instructor-in-the-loop feedback systems \cite{tang2025sphere,kaputa2026simstep}. Finally, whether this kind of modality-matched delivery produces measurably different learning outcomes compared to fixed-modality baselines is an empirical question that the learning sciences community should investigate.

\vspace{-4pt}
\section{Related Work}
\vspace{-4pt}

\paragraph{Generative UI for Education.}
GenUI is considered in the HCI literature as an approach in which LLMs can be used to generate interface components rather than rendering pre-authored layouts, adapting visually and functionally to user context \cite{pott2026taxonomy}. In EdTech, this prior works have started to implement this idea in user-centric systems. For example, MAIC-UI \cite{tu2026maic} enables educators to generate interactive courseware from uploaded documents without coding, or SimStep \cite{kaputa2026simstep} helps the educators in the process of authoring generated simulations.

\paragraph{Multimodal Content Generation.}
Another research direction has explored whether LLMs actually have the ability to generate high-quality multimodal educational content. EduIllustrate \cite{bi2026eduillustrate} proposes a benchmark for evaluating LLM-generated text-diagram explanations for K-12 STEM and finds notable variation across models on dimensions such as visual consistency and pedagogical accuracy. MAGMA-Edu \cite{wu2025magma} presents a multi-agent framework for generating coordinated text and diagram content for education, as most prior approaches are text-only, despite education being inherently multimodal. The Universal Design for Learning framework \cite{rose2000universal} similarly frames representation diversity, including multiple means of engagement, representation, and action, as a design-time principle rather than a runtime accommodation. However, these systems are mostly research prototypes evaluated on content quality, not tools that educators can use to author accessible learning material and deploy them at scale.

\paragraph{Adaptive Authoring and Instructor Control.}
Learning analytics and AIED communities have produced extensive research on instructor-facing tools that operate at delivery time, involving instructors in the education loop. For example, VizGroup \cite{tang2024vizgroup} and ClassAid \cite{zhang2026classaid} provide real-time dashboards and orchestration capabilities for instructors to monitor and respond to student behavior during live sessions, or SPHERE \cite{tang2025sphere} addresses the challenge of scaling personalized feedback in programming education by combining LLM generated outputs with a structured instructor review process, a ``strategy-detail-verify'' approach that keeps educators in the loop before feedback reaches students. Together, prior works show that A) instructor verification and oversight of AI-generated content before delivery is both feasible and crucial for maintaining quality, and B) the interface for that oversight must be designed carefully to be usable by the instructor at scale. However, these systems mainly focus on feedback or analytics provided on top, not focusing on generating content with AI; this is while co-design studies with educators have shown a low trust in AI-generated content and a strong preference for maintaining more control over AI \cite{leanlab2024humans}.

\paragraph{Equitable Delivery and Low-Bandwidth Contexts.}
The Learning at Scale community has considered equity and scale as central concerns, examining how learning technology can reach learners beyond well-resourced, high-connectivity environments. Studies of WhatsApp-based learning in low- and middle-income countries \cite{jordan2023can} show the feasibility of, and the demand for, low-bandwidth educational content delivery. Those who benefit most from scalable EdTech are often those for whom current systems are least accessible; thus, a paradigm that generates only rich, interactive, and bandwidth-intensive representations at runtime cannot necessarily reach them.

\vspace{-4pt}
\section{The Card-Based GenUI Paradigm}
\vspace{-4pt}

\begin{figure*}[t]
    \centering
    \includegraphics[width=0.9\textwidth]{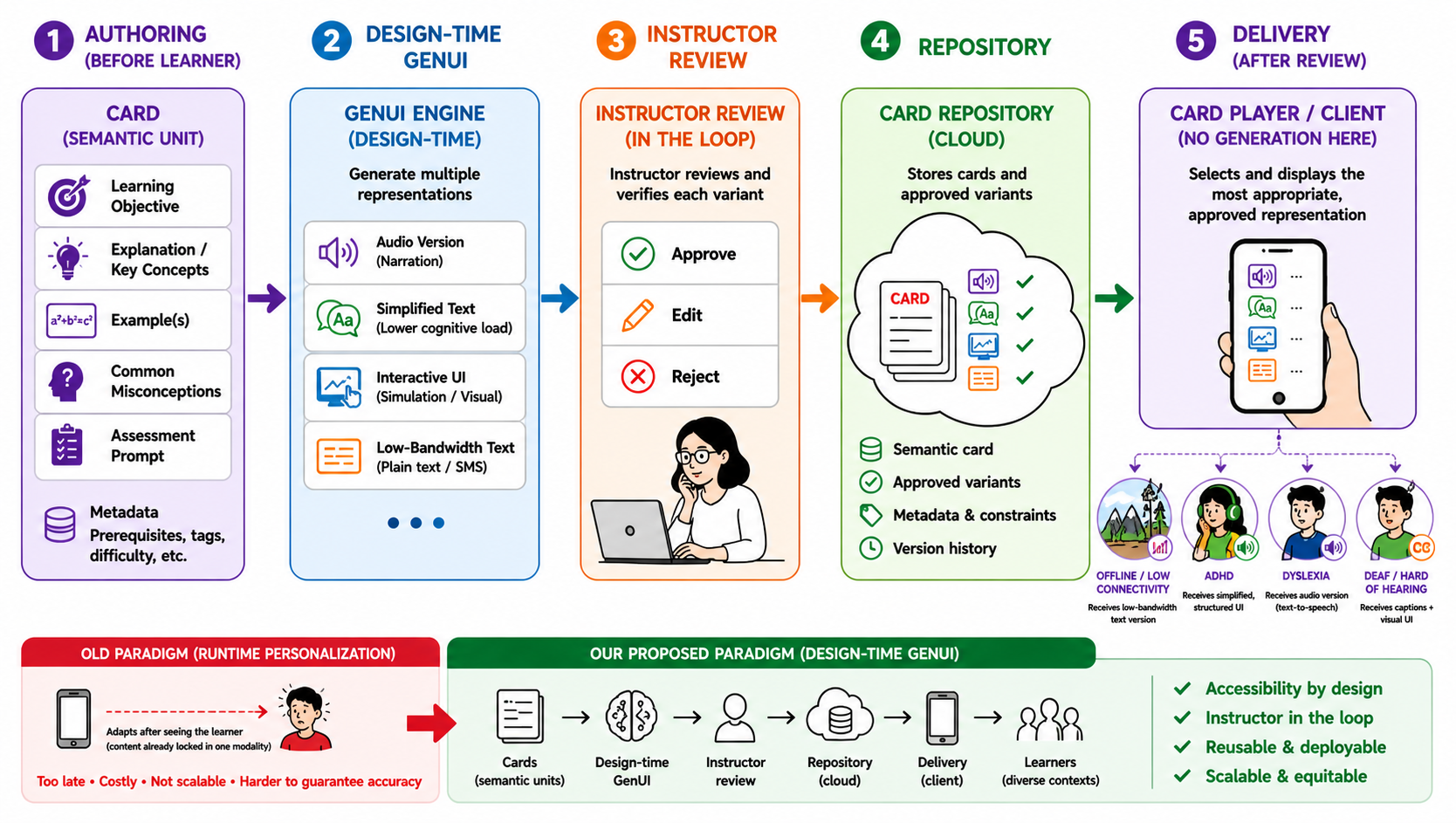}
    \vspace{-8pt}
    \caption{
    Overview of our proposed card-based design-time GenUI paradigm (generated with the \textit{ChatGPT Image Generation model} as an artifact in the end of a focus group discussion we held on our proposal). Educational content is authored as modality-agnostic cards, transformed into multiple representations at design time, reviewed by instructors, and delivered as pre-generated variants to support diverse learner needs without runtime generation.}
    \label{fig:genui_pipeline}
    \vspace{-15pt}
\end{figure*}

We propose a card-based GenUI paradigm, illustrated in Fig.~\ref{fig:genui_pipeline}, in which educational content is authored once as structured semantic material, and then transformed into multiple interface representations at the \textit{learning design} time. The figure synthesizes insights from our focus group discussions and was generated with the assistance of generative AI. We treat the learning unit, and not the specific interface, as the primary authored object. In this view, an educator does not begin by designing a webpage, conversation flow, slide, or simulation. Instead, the educator defines a bounded learning unit as a \textit{card}, and the system generates possible representations of that unit for different learners, contexts, and constraints.

\subsection{Cards as Semantic Content Units}

A card is a bounded knowledge unit: the smallest meaningful unit of instructional content that can be authored, reviewed, sequenced, and delivered independently. A card may introduce a concept, explain a procedure, present a worked example, or provide practice. What makes it a card is not its visual appearance, but its semantic structure.

Each card encodes \textit{what} the learner is intended to understand, separately from \textit{how} that understanding will be represented. For example, a card on the quadratic formula would include the learning objective, prerequisite concepts, key explanation, notation, examples, common misconceptions, and perhaps a short assessment prompt. It would not initially specify whether the learner should encounter this material as an interactive graph, an audio explanation, a simplified textual explanation, or a low-bandwidth message. Those are representations of the card, not the card itself.
This separation is important because educational interfaces often conflate content and representation. A lesson authored as a slide deck inherits the assumptions of slides: visual layout, sequential navigation, and necessitating relatively high bandwidth. Similarly, a lesson authored as a webpage inherits the assumptions of web interaction, and a lesson authored as a video assumes audio-visual access and sufficient connectivity. Once content is created inside one form, adapting it to another form becomes a translation problem. The card-based paradigm instead asks educators to author at the level of learning intent before representational commitments are made.

A card can therefore be understood as a structured object containing fields such as learning objective, concept description, examples, prerequisite knowledge, constraints, assessment item, and source material. These fields bound the generation; rather than asking a model to generate arbitrary educational content, the system asks it to generate representations of a specific, instructor-authored unit. This boundary is central to our approach: cards make generation granular enough to review and reuse, but keeps it meaningful enough to stay pedagogical coherent.

Cards also support composition: a course can be represented as a sequence or graph of cards, where educators define prerequisite relationships, optional branches, remediation paths, or thematic clusters. This creates the basis for a \textit{card repository}: a reusable collection of semantically structured learning units and their approved representation variants. Such repositories could support sharing, remixing, localization, and deployment across different platforms, while preserving instructor control over the underlying learning content.

\subsection{Learning Design-Time GenUI}

We suggest applying GenUI at authoring time, before the card is delivered to learners. Once an educator creates a card, a GenAI engine generates a defined set of representation variants. These variants may include, for example, an interactive simulation, an audio-first narration that doesn't assume that the learner can see equations or diagrams, a simplified text explanation to reduce unnecessary cognitive load without removing essential conceptual structure, or a low-bandwidth text version communicable over WhatsApp, depending on the domain, target group, and deployment constraints.

This learning design-time approach differs from runtime personalization. In runtime GenUI, the system generates or adapts interface components while a learner is interacting with the system. This can be valuable, especially for hints, feedback, or adaptive scaffolding. In contrast, learning design-time GenUI generates representations once, stores variants approved by the instructor, and delivers them later through appropriate clients. This distinction is particularly important for educational accuracy: in runtime generation, errors may be exposed directly to learners, but if the same representation is generated during authoring, the instructor can review and correct it before deployment. The original author is the best person to verify whether the generated variant preserves the intended concept, uses appropriate terminology, avoids misleading simplifications, and most importantly, aligns with the pedagogical purpose of the card.

Learning design-time generation also changes the cost structure. Once approved, a card representation can be delivered without repeatedly invoking an LLM. This is important for large courses, high-enrollment systems, and low-resource settings where per-learner inference costs may be prohibitive. The client may still invoke AI in runtime for selected functions, such as answering questions or generating feedback, but the core representational diversity of the learning material would not depend on continuous model calls.

\subsection{Educator Agency and Review}

The card-based paradigm depends on educator agency. Educators would be able to preview and edit representations. The system therefore remains an authoring assistant, not an autonomous teacher. This is the mechanism through which this paradigm addresses trust and accuracy: for each card, the system would surface the generated variant alongside the original semantic fields and ask the educator to verify whether the representation preserves the learning objective, introduces any factual or conceptual errors, and remains appropriate for the target learners.

One possible review structure is inspired from strategy-detail-verify workflows. For strategy, the system explains why a particular representation was generated: e.g., a low-bandwidth text card was optimized for weak cellular data access, or an interactive simulation was generated to support manipulation of variables. At the detail level, the educator inspects the specific generated content. At the verification level, the educator approves, edits, or rejects the variant.

Finally, once cards and their representation variants are approved, they can be stored in a card repository. A repository would contain the semantic card, metadata, approved variants, version history, and deployment constraints. Different clients could then render the same card repository in different ways; for example, a web client might display the interactive version, a mobile-first client might prioritize simplified text and audio, and a WhatsApp-based client might deliver the low-bandwidth representation.

\vspace{-4pt}
\section{Implications for AI, HCI, and Learning Sciences}
\vspace{-4pt}
We outline a set of key considerations for each community.

\textbf{Implications for AI.} Hallucination and error risks are not uniform across representations. Errors in text or visuals may be more easily identified and corrected, while errors in audio narration or interactive components may be less transparent to learners. This suggests the need for representation-specific validation strategies, rather than treating all generated outputs uniformly. We propose to partly address this challenge by design-time instructor review, but systematic AI evaluation for detecting errors across modalities remain crucial.

\textbf{Implications for HCI.} In this paradigm, each card would have multiple valid representations, raising questions about how to evaluate usability, consistency, and quality across variants. Traditional evaluation methods assume a fixed interface; in contrast, this paradigm would require assessing families of interfaces derived from the same underlying content.
Also, reviewing multiple generated variants per card introduces a new interaction problem: how to make this process efficient, interpretable, and trustworthy for non-technical instructors. While prior research, such as the strategy-detail-verify workflow \cite{tang2025sphere}, suggest directions for structuring this interaction, dedicated interface design for multi-representation authoring remains largely unexplored. Finally, an open question remains on learner control: if multiple representations of the same card exist, should learners choose their preferred modality, or should the system assign representations based on inferred needs or context? 

\textbf{Implications for Learning Sciences.} This paradigm introduces empirical questions about how various representation affects learning differently. The assumption that card variants are interchangeable from a learning perspective is an empirical claim that requires validation.
In addition, the card structure aligns naturally with existing instructional strategies such as scaffolding, and modular sequencing. However, how card granularity and sequencing interact with these strategies remains an open question. Finally, the paradigm raises questions about equity. Prior work has shown that interactive and adaptive systems can disproportionately benefit lower-performing learners \cite{tu2026maic}; it is not clear whether providing modality-matched representations amplifies or mitigates such effects. For example, while offering simplified or audio-first representations may support accessibility, it could also influence how learners engage with material and affect learning outcomes. Understanding these specific dynamics is essential for evaluating the broader impact of the approach.

\vspace{-4pt}
\section{Conclusion}
\vspace{-4pt}
We argue that the current emphasis on runtime personalization in GenUI EdTech overlooks a more fundamental opportunity at the authoring layer. We proposed a card-based GenUI paradigm in which learning content is structured as semantic units and multiple representations are generated and verified at design time. This addresses limitations in cost, accuracy, and deployability. We hope this perspective encourages the community to reconsider GenUI use in EdTech and to explore authoring-time generation as complementary to runtime adaptation.

\begin{credits}
\vspace{-6pt}
\subsubsection{\ackname} Thanks to ChatGPT and Claude for co-authoring the text through improving format and structure, as well as suggestions for change and addition.
\end{credits}
\bibliographystyle{splncs04}
\vspace{-8pt}
\bibliography{reflectiontool}
\vspace{-8pt}

\end{document}